\documentclass[twocolumn,showpacs,preprintnumbers,amsmath,amssymb]{revtex4}
\usepackage{graphicx}% Include figure files
\usepackage{dcolumn}% Align table columns on decimal point
\usepackage{bm}% bold math
\newcommand{\eq}{\begin{equation}}
\newcommand{\qe}{\end{equation}}

\begin{document}

\title{\textit{In Situ} Treatment of a Scanning Gate Microscopy Tip}

\author{A. E. Gildemeister}
\email{gildemeister@phys.ethz.ch}
\author{T. Ihn}
\author{M. Sigrist}
\author{K. Ensslin}

\affiliation{Laboratory of Solid State Physics, ETH Z{\"u}rich,
CH-8093 Z{\"u}rich, Switzerland}

\author{D. C. Driscoll}
\author{A. C. Gossard}

\affiliation{Materials Department, University of California,
Santa Barbara, CA-93106, USA}

\begin{abstract}
In scanning gate microscopy, where the tip of a scanning force microscope is used as a movable gate to study electronic transport in nanostructures, the shape and magnitude of the tip-induced potential are important for the resolution and interpretation of the measurements. Contaminations picked up during topography scans may significantly alter this potential. We present an \textit{in situ} high-field treatment of the tip that improves the tip-induced potential. A quantum dot was used to measure the tip-induced potential.
\end{abstract}

\pacs{07.79.-v, 73.21.La, 73.23.Hk}

\maketitle

In scanning gate microscopy (SGM) the sharp conducting tip of a scanning force microscope (SFM) is used as a movable gate to study electrical transport with high spatial resolution. This technique has been  applied to study the classical \cite{Baumgartner:2006} and quantum Hall effect \cite{Tessmer:1998, Steele:2005}, quantum point contacts \cite{Topinka:2001,Crook:2006,Cunha:2006,Pioda:2007,Gildemeister2:2007}, and quantum dots \cite{Woodside:2002,Pioda:2004,Fallahi:2005,Gildemeister3:2007,Kicin:2005}. 
Because in SGM the sample interacts capacitively with the tip of the microscope it is important to understand and control the potential the tip induces in the sample. It is empirically known that the shape of the tip-induced potential may be unexpectedly complex \cite{Finkelstein:2000, Pioda:2007,Kicin:2005}. Recently, we have found that indeed the tip-induced potential may have two components, one that depends on the tip bias and one that is independent of tip bias \cite{Gildemeister3:2007,Gildemeister2:2007,Kicin:2005}. This complicates the interpretation of SGM measurements.

Here we present a method that improves the tip-induced potential by simplifying its geometry. Our working hypothesis is that charged dielectric particles clinging to the tip create the part of the tip-induced potential that is independent of the tip bias \cite{Finkelstein:2000,Gildemeister3:2007}. Removing these particles would leave us with a tip-induced potential that depends only on the shape of the conducting tip and can be controlled more easily. 

It is difficult to prevent contamination with particles for four reasons. First, the samples studied by SGM are typically patterned with complex nanostructures which, once they are finished, easily get damaged by thorough surface cleaning procedures. Second, the experiments are mostly performed at very low temperatures and it is usually necessary to make large topography scans in order to locate the structure of interest which has shifted laterally relative to the tip while cooling down the microscope. This increases the probability of picking up a particle. Third, the tips are usually prepared ex situ and transferred to the microscope under ambient conditions. Replacing a contaminated tip is cumbersome because starting a new cooldown with a fresh tip is time-consuming. Fourth, SGM is usually done in a decent vacuum of $p<10^{-5}$~mbar but not in ultra high vacuum.

In scanning tunneling microscopy (STM) it is common practice to clean tips by suddenly raising the tip bias to about 10~V, a procedure known as high-field treatment or ``flashing" \cite{Wintterlin:1989,Chen:1993}. We have adapted this technique for use in SGM. 

It is important to stress a difference between STM and SGM. In STM the tip is usually closer than 1~nm to the surface and for the highest resolution images the tunnel current passes through the single foremost atom of the tip. Because the tunnel current decreases exponentially with distance, additional tips or impurities on the tip that are set back by more than a few atomic layers do not affect the measurement. In SGM the distance between tip and sample needs to be larger, typically on the order of tens or hundreds of nanometers, because the surface is usually not flat but lithographically patterned. The tip is coupled capacitively to the sample and we need to take the long-range nature of electrostatic interaction into account. Therefore the exact shape of the tip and contaminations do influence the measurement, even when they are significantly set back from the foremost tip. 

\begin{figure*}
\includegraphics[width=16.4cm]{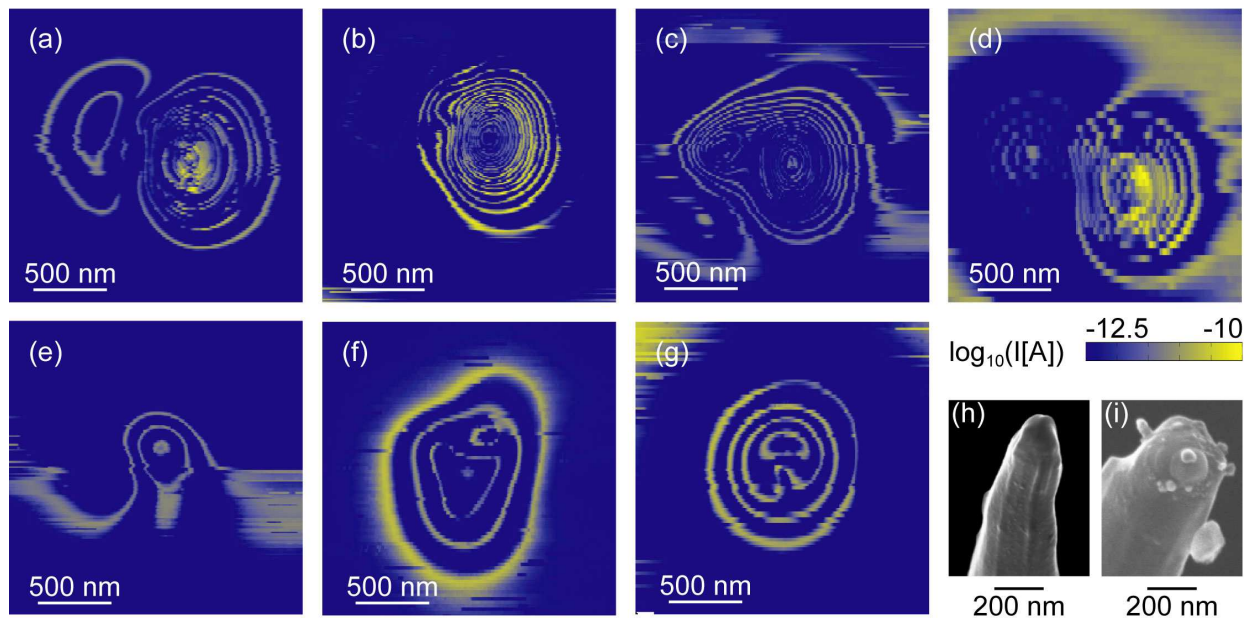}
\caption{\label{fig:flash} (Color online) (a-g) Scanning gate measurements that show the current through a quantum dot as a function of tip position. Lines of high current are equipotential lines of the tip-induced potential. (a) Shows the tip-induced potential recorded after first cooling down the microscope. In (b-g) we can see how the tip-induced potential changed due to the high-field treatment. We also show SEM images of the tip before (i) and after (h) the measurement cycle.}
\end{figure*}

We have used a quantum dot to measure the tip-induced potential. The dot was defined by local anodic oxidation  on a GaAs/AlGaAs heterostructure with a two-dimensional electron gas residing 34~nm below the surface. A thin film of Ti was evaporated on the surface and again oxidized to form top gates \cite{Sigrist:2004}. The area above the dot was completely oxidized to allow capacitive coupling between tip and dot only. The measurements were performed in a dilution refrigerator cooled SFM \cite{Gildemeister1:2007} at an electronic temperature of about 190~mK. The tip was glued on a tuning fork sensor and the setup allowed both STM operation and SFM operation. For the latter the change of the tuning fork sensor's resonance frequency was used to detect the force between tip and sample.

The PtIr tip was electrochemically etched from a 15~$\mu$m wire at room temperature. For etching we used a solution of 7~g CaCl$_2$ $\cdot$ 2H$_2$O in 40~ml H$_2$O and 2~ml acetone \cite{Rogers:2000}. We applied an ac voltage of 7~$V_{\textrm{pp}}$ at 700~Hz between the wire and a small gold ring around it that held a droplet of the etchant. The voltage was applied until the wire fell apart to form the tip.

The quantum dot was tuned to the Coulomb blockade regime where its electrical conductance is very low unless one of the dot's quantized energy levels comes into resonance with the chemical potential in the source and drain leads. We applied a small ac bias of 20~$\mu$V across the dot and mapped the current through the dot as the tip of the microscope was scanned over it. Figures \ref{fig:flash}(a-g) show seven such measurements where the tip was scanned at a constant height of about 100~nm over the surface. In between two images the cleaning procedure described below was performed. At most tip positions the current is low and only when the tip-induced potential aligns an energy level of the dot with the Fermi level of the leads we see an enhanced current. The lines of higher current are equipotential lines of the tip-induced potential \cite{Gildemeister3:2007}. At every line the number of electrons on the dot is changed by one and the energy separation between two subsequent lines is roughly the charging energy of the dot, here about 1~meV.

In Fig. \ref{fig:flash}(h) wee see a scanning electron microscope (SEM) image of the tip taken before the microscope was cooled down. It is clean and the foremost tip has a diameter of less than 100~nm. After cooling down the microscope we had to make several topography scans in order to locate the quantum dot. Figure \ref{fig:flash}(i) shows how after the cooldown the tip was deformed and contaminated with additional particles compared to Fig. \ref{fig:flash}(h).

In Fig. \ref{fig:flash}(a) we see the potential of the grounded tip when it was first scanned over the dot. Clearly, a double tip had developed \cite{Pioda:2007} and we see two extrema some 500~nm apart. One has a magnitude of about 2~meV while the other has a magnitude of more than 10~meV as we can estimate by counting the equipotential lines. Such a potential is typically suboptimal to be used for SGM investigations because it will lead to a poor resolution and complicate the interpretation \cite{Kicin:2005}.

In order to improve the tip potential we have now tried different cleaning procedures. In each case we used our coarse positioning system to move the tip over one of the metallic leads that connect to the Ti film about 10~$\mu$m away from the quantum dot. We moved to the thicker lead because it withstands high currents better than the thin Ti film. In STM mode we moved the tip to the surface, switched to a higher tip bias and withdrew the tip again. At high tip bias we temporarily observed currents of over 100~nA between tip and surface and the sample temperature rose by about 25~mK. In order to check the result we used the coarse positioning system again to move the tip over the dot. This required some topography scanning to verify the position. With the dot we could then map the tip-induced potential again. The cleaning procedure influenced the sample and some fine-tuning was necessary to bring the quantum dot back into the Coulomb blockade regime. This fine-tuning has a negligible influence on the images presented in Fig. \ref{fig:flash}(a-g).

In Table~\ref{tab:flashs} we list the parameters that were used in the different repetitions of the high-field treatment. We applied a low bias voltage $V_l$ to the tip, brought it close to the surface until the current setpoint $I$ was reached and then applied the higher bias $V_h$. In each step this was repeated a few times.

\begin{table}
\caption{\label{tab:flashs} Parameters used for the successive high-field treatments of the tip. $V_l$ is the tip bias used in STM mode with a current setpoint $I$ while $V_h$ is the tip bias used for cleaning.}
\begin{ruledtabular}
\begin{tabular}{lllll}
No. & Result & $V_l$ [V] & $I$ [nA] & $V_h$ [V] \\
\hline
1 & Fig. \ref{fig:flash}(b) & +1.0 & 0.2 & +5  \\
2 & Fig. \ref{fig:flash}(c) & +1.5 & 0.2 & +5  \\
3 & Fig. \ref{fig:flash}(d) & -1.5 & 0.2 & -8  \\
4 & Fig. \ref{fig:flash}(e) & -1.5 & 2.0 & -10 \\
5 & Fig. \ref{fig:flash}(f) & -1.5 & 1.0 & -10 \\
6 & Fig. \ref{fig:flash}(g) & -1.5 & 1.0 & -10 \\
\end{tabular}
\end{ruledtabular}
\end{table}

Figures \ref{fig:flash}(b-g) show the resulting scanning gate images which resemble the tip-induced potential. The tip was grounded except in Fig. \ref{fig:flash}(f) where it was biased to +100~mV. The tip-induced potential  changed significantly, both in shape and magnitude. In the first four images we see closely packed equipotential lines, indicative of a relatively steep potential. In the remaining three images the equipotential lines are further apart and the potential is flatter. There is still more than one extremum but the extrema are closer to each other. This could indicate that a highly charged particle has fallen off the tip. What remains is the potential of the metallic tip, possibly decorated by a particle charged less than before. The  potential shown last in Fig. \ref{fig:flash}(g) is, albeit still imperfect, much simpler than the one in Fig. \ref{fig:flash}(a). The magnitude decreased from more than 10~meV to about 4~meV and the shape became more symmetric. We did not find that any particular set of cleaning parameters led to better results than others. In order to locate the quantum dot, we had to make a topography scan between the cleaning procedure and the measurement of the tip-induced potential each time. This unavoidable scan may have contaminated the tip anew. It seems, for example, that after the first cleaning a particle was removed and after the second cleaning a new particle was picked up (Figs. \ref{fig:flash}(a-c)). Two cleaning procedures later (Fig. \ref{fig:flash}(e)) this particle is removed again.

We found that, before the high-field treatment, the $z$-position at which the tip encountered the surface was several tens of nanometers closer to the surface when measured in STM mode  compared  to what we measured by mechanically oscillating the tip. This could be due to an insulating particle on the tip. The observation of a tip bias independent potential and the changes of the tip-induced potential due to the high-field treatment also indicate the presence of charged insulating particles on the tip. This makes us confident that a contamination of the tip is the reason for odd tip-induced potentials. In Fig. \ref{fig:flash}(i) we show an SEM picture of the contaminated tip after it had been used in measurements for several weeks. It is unclear, however,  when exactly these particles were picked up.

In conclusion, we have shown how a high-field treatment can be used to improve the tip-induced potential of SGM probes. Additional measures to clean the surface and to avoid topography scans during which the tip comes very close to the surface will be necessary to fully control the tip-induced potential.

We would like to thank U. Ramsperger and J. Repp for discussions and P. W{\"a}gli for making the SEM images. Financial support from ETH Z{\"u}rich is gratefully acknowledged.

%%%%%%%%%%%%%%%%%%%%%%%%%%%%%%%%%%%%%%%%%%%%%%%%%%%%%%%%%%%

%\bibliography{bib008}% Produces the bibliography via BibTeX.

\end{document}